# Modeling threshold exceedance probabilities of spatially correlated time series

**Dana Draghicescu**[*]

*Department of Mathematics and Statistics*
*Hunter College of the City University of New York, USA*
*e-mail:* dana.draghicescu@hunter.cuny.edu

**Rosaria Ignaccolo**

*Dipartimento di Statistica e Matematica Applicata*
*Università degli Studi di Torino, Italy*
*e-mail:* ignaccolo@econ.unito.it

**Abstract:** The Commission of the European Union, as well the United States Environmental Protection Agency, have set limit values for some pollutants in the ambient air that have been shown to have adverse effects on human and environmental health. It is therefore important to identify regions where the probability of exceeding those limits is high. We propose a two-step procedure for estimating the probability of exceeding the legal limits that combines smoothing in the time domain with spatial interpolation. For illustration, we show an application to particulate matter with diameter less than 10 microns ($PM_{10}$) in the North-Italian region Piemonte.

**AMS 2000 subject classifications:** primary 62-09; secondary 62G99.

**Keywords and phrases:** threshold exceedance probability, $PM_{10}$, smoothing, spatial interpolation, spatial time series, visualization.

## 1. Introduction

For many physical processes, it is important to model the probability of exceeding specific thresholds, above which negative effects on human health and the environment have been observed. Examples include climatological processes (temperature, precipitation), and air pollution processes (ozone, fine particles).

One way of estimating exceedances over high thresholds is given by methodology using extreme value theory. A review of statistical techniques for analyzing extreme values can be found in Smith (1989), together with a detailed analysis of ozone time series. Davison and Smith (1990) analyze the generalized Pareto distribution used for modeling exceedances over high thresholds, and give applications to river flows and wave heights. Tawn (1992) uses a modified version of the joint probabilities method for the analysis of extremes of non-stationary sequences, with an illustration on estimation of extreme sea levels exceeded by the annual maximum hourly sea height with a prescribed probability. Niu (1997) discusses extremal properties of non-stationary time series, and shows an application to stratospheric ozone in Chicago, concerning the exceedance of daily maxima over the national standard. Piegorsch et al. (1998) present a thorough overview of statistical methodology for extreme values and rare events with focus on environmental applications. Chavez-Demoulin and Embrechts (2004) describe smooth non-stationary generalized additive models for extremes, and incorporate splines in a model for exceedances over high thresholds applied to finance and insurance problems.

---

[*]Corresponding author, phone 212-772-5748, fax 212-772-4858

An alternative, more general approach for modeling exceedances, is via methodology in time series and spatial statistics for probability distribution functions and quantiles. Also, it is often the case that environmental data have large temporal coverage (hourly or daily measurements over long periods), and smaller spatial coverage (a relatively small number of monitoring locations). Such data can be viewed as a collection of long time series that are spatially correlated. The problem of interest is to map the exceedance probabilities over a fixed threshold. Many times, the observed time series display different patterns at the different monitoring sites, and show significant departures from stationarity and Gaussianity.

In this note we introduce a two-stage exploratory technique that combines smoothing in the time domain with spatial interpolation, to produce maps of exceedance probabilities for the spatial domain of interest. Motivated by the good temporal coverage of the data, and in order to provide a flexible, comprehensive approach, we start by smoothing the observed 0 or 1 exceedance probabilities, then interpolate the resulting estimates by using a parametric spatial covariance function.

A flexible way to characterize complex temporal dependence, is via a time-varying transformation $G(t, Z_t)$ of a stationary process $Z_t$. The physical evolution of many real-life processes makes this a plausible working assumption. As the unknown transformation $G$ is allowed to vary with time, the probability distribution function of the resulting process may also change, and therefore the process may not be stationary. Ghosh et al. (1997) studied the asymptotic properties of a nonparametric estimator of the marginal probability distribution function in this setting, where the underlying process $Z_t$ was assumed to be Gaussian and having long memory. A similar estimator was analyzed in Draghicescu (2003), Draghicescu and Ghosh (2003) for the case when the underlying Gaussian process has short memory (under the general assumption that the correlations are summable). A data-driven procedure for optimal bandwidth selection for these kernel estimators was proposed in Ghosh and Draghicescu (2002), and discussed in detail in Draghicescu (2003). A new kernel distribution function estimator was recently discussed in Swanepoel and Van Graan (2005), where a data-based choice of bandwidth was also proposed. The method holds for independent, identically distributed (*iid*) data, and can be extended for weakly dependent observations. Bosq (1998) provides a comprehensive overview of nonparametric methods for stochastic processes. However, there are many open problems regarding spatial modeling of distribution functions. One immediate, albeit naive approach, is given by the so-called *indicator kriging* (Chilès and Delfiner 1999, page 383), which is an adaptation of universal kriging (spatial interpolation). Recently Short et al. (2005) introduced a fully hierarchical approach for modeling distribution functions for bivariate spatial processes, using a Bayesian framework implemented via MCMC methods.

The Commission of the European Union, as well as the United States Environmental Protection Agency, have set limit values for some pollutants in the ambient air, that were proved to have a negative impact on human and environmental health. In particular, recent studies linked traffic-related pollutants to increased risks of morbidity and mortality due to respiratory and cardiovascular illness (see for example Samet et al. 2000 and the references therein). It is therefore important to identify regions where the probability of exceeding these legal limits is high.

In this paper, we focus on particulate matter with diameter less than 10 microns ($PM_{10}$). Exploratory analyses and basic statistical models for this pollutant are employed in McKendry (2000), Ignaccolo and Nicolis (2005), Rajsic et al. (2004), among others. A review of recent studies on particulate matter is given in Schimek (2003), where a semi-parametric model including weather information is used to link particulate matter to hospital admissions in a regional study, controlling for potential spatial dependencies. In contrast to these studies, where $PM_{10}$ is modeled directly, we focus on the probability of exceeding the legal limit. We use a two-stage procedure to estimate the space-time exceedance probability over a given threshold. The paper is organized as follows. The data set that motivated this study is described in Section 2. Section 3 is devoted to statistical methodology, followed by Monte Carlo simulations in Section 4, and an application to air pollution data in Section 5. A brief discussion is given is Section 6.

## 2. A motivating data set

We analyze daily $PM_{10}$ concentrations (in $\mu g/m^3$) during 2004 at 22 sites in the North-Italian region Piemonte. The data were collected through the information system *AriaWeb Regione Piemonte*. A detailed description of this monitoring network can be found in Ignaccolo and Nicolis (2005). The 22 records used in this study were selected from Low Volume Gravimetric (LVG) monitors, such that the amount of missing data did not exceed 10%. The missing values were imputed by using kernel regression smoothing with adaptive plug-in bandwidth (Gasser et al. 1991). The maximum allowable number of days to exceed 50 $\mu g/m^3$ is 35, and therefore 50 $\mu g/m^3$ is the threshold corresponding to the 0.904 quantile during one year. Detailed explanations of the European norms for air pollution are given in van Aalst et al. (1998). Figure 1 shows the locations of these 22 sites, together with the respective 0.904 quantiles.

It can be observed that the sites near the Alps have lower $PM_{10}$ values, whereas higher pollution levels are detected in the valleys, closer to urban areas. Regarding the temporal variations of this pollutant, we found that the behavior of $PM_{10}$ has different patterns at different locations, yielding different patterns in the behavior of the probability of exceeding 50 $\mu g/m^3$. Given the good temporal coverage of the data, we employ smoothing in the time domain to first model these exceedance probabilities in a flexible and comprehensive way. For an illustration, we show in Figure 2 the series of daily $PM_{10}$ concentrations for year 2004 at three locations, the corresponding 0 or 1 probabilities of exceeding 50 $\mu g/m^3$, and the associated smoothed exceedance probabilities.

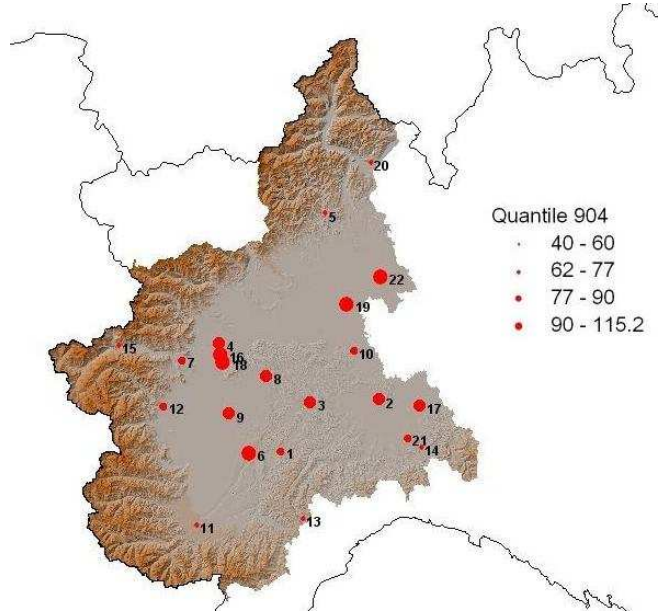

FIG 1. *Locations of 22 air pollution monitoring stations in Piemonte; superimposed 2004 annual 0.904 quantiles of daily $PM_{10}$ ($\mu g/m^3$).*

With respect to the temporal dependence structure, these 22 time series displayed short-range dependence, as shown by the boxplots of their autocorrelation functions in Figure 3.

The same data set is used in Bande, Ignaccolo and Nicolis (2006), where additional meteorological and geographical information is included in a model for the space-time $PM_{10}$ trend (mean function). With regard to modeling the probability of exceeding the cutoff $PM_{10}$ legal value, the preliminary study Draghicescu and Ignaccolo (2005) proposes a sequential modeling technique for the maps of exceedance probabilities in the same region, using data from 17 monitors for the year 2003. This method will be briefly described in Section 4, and used for comparisons in the simulations and applications.

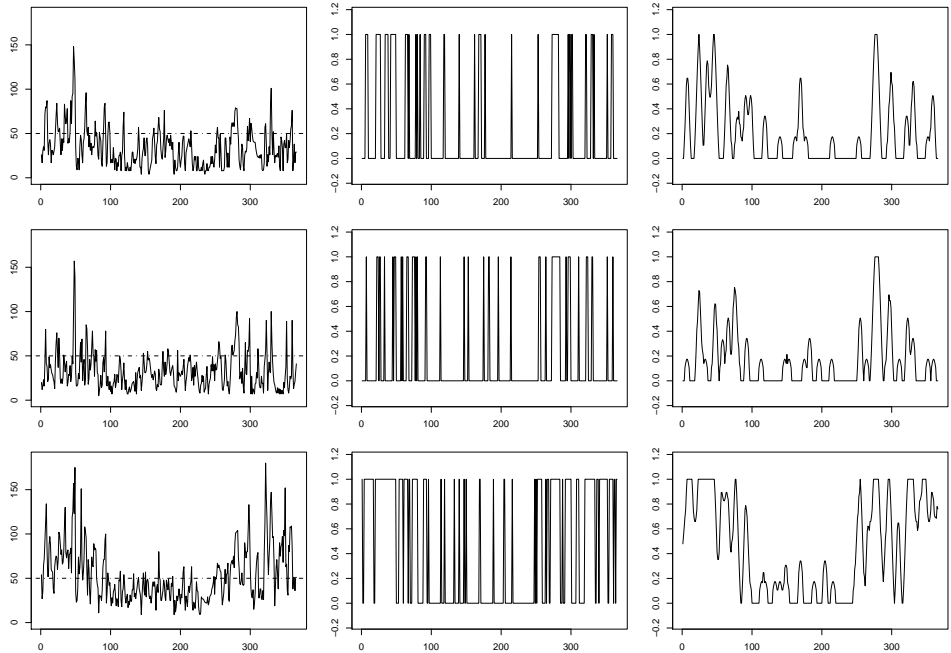

Fig 2. *Time series of daily $PM_{10}$ concentrations, year 2004 (left), corresponding 0 or 1 series representing the 50 µg/$m^3$ exceedance probabilities (middle), and smoothed exceedance probabilities (right) at station 5 (top), station 19 (middle), and station 11, (bottom).*

## 3. Theoretical framework

Assume that at each location $s \in D$ (for some domain $D \in \mathbf{R}^2$) we observe a temporal process of the form $X_s(t) = G_s(t, Z_s(t))$, where $G_s$ is an unknown transformation, $Z_s$ is a standardized stationary Gaussian process with $\gamma_s(l) := cov(Z_s(t), Z_s(t+l))$, such that $\sum_{l=-\infty}^{\infty} |\gamma_s(l)| < \infty$. This general class of processes includes non-stationary and non-Gaussian situations, and is therefore suitable to model complex environmental data sets. Also note that no parametric assumptions are made on the temporal covariance structure of the process. For fixed $x_0 \in \mathbf{R}$, define the exceedance probability $\mathbf{P}_{x_0}(t,s) = P(X_s(t) \geq x_0)$. By using the axioms of probability, it is immediate to see that $\mathbf{P}_{x_0}(t,s)$ takes values in $[0,1]$ and is non-increasing in $x_0$. The problem of interest is to predict $\mathbf{P}_{x_0}(t, s^*)$ at location $s^* \in D$ where there are no observations, and at any time $t$, based on observations of the process $X_s(t)$ at $n$ time points $t_1, \ldots, t_n$, and at $m$ spatial locations $s_1, \ldots, s_m$. Typically $m$ is much smaller than $n$. Furthermore, for fixed $t$, these exceedance probabilities are assumed to be isotropic in space, meaning that their spatial covariances depend only on the Euclidean distance between the respective locations.

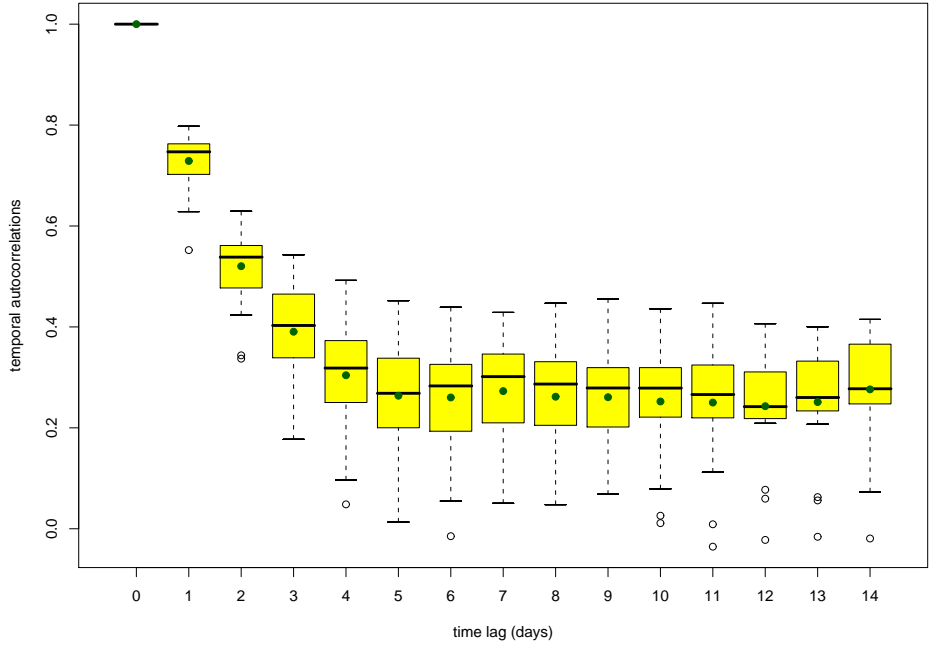

Fig 3. *Boxplots (over the 22 locations) of the empirical autocorrelation functions for the time series of daily $PM_{10}$ at these locations, year 2004.*

The method we propose consists of two steps. We start by modeling the temporal exceedance probabilities for any location where the process is observed. Let $X_s(t_1), \ldots, X_s(t_n)$ be the daily $PM_{10}$ observed concentrations, assumed to be realizations of a process $X_s(t) = G_s(t, Z_s(t))$, and $x_0$ the $PM_{10}$ legal limit. We stress again that the true exceedance probability may change with time and space. Therefore the empirical estimator $\mathbf{P}_{n,m}(x_0) := \frac{1}{nm} \sum_{i=1}^{n} \sum_{j=1}^{m} 1_{\{X_{s_j}(t_i) \geq x_0\}}$ (analogous to the empirical distribution function) is too rough, and cannot capture such changes, making it necessary to consider different weights, that should depend on the local behavior of the process. As the problem of interest is not to assess the mean function of the process, but the exceedance probability over the given threshold $x_0$, we focus on the indicator process $1_{\{X_s(t) \geq x_0\}}$. We assume further that the changes of this indicator process with time are smooth for all $s \in D$, namely that $1_{\{X_s(t_i) \geq x_0\}} - \mathbf{P}_{x_0}(t_i, s) = \sum_{k=1}^{\infty} c_{k,s,x_0}(t_i) H_k(Z_s(t_i))$, where $t_i = \frac{i}{n}$ are rescaled time points, $H_k$ denotes the Hermite polynomial of degree $k$, and the coefficients $c_{k,s,x_0}$ are twice continuously differentiable with respect to $t$, and continuous with respect to $s$ and $x_0$. For examples of time-varying transformations of Gaussian processes and detailed explanations of assumptions similar to the ones above, we refer to Draghicescu (2003). In the second step we use spatial interpolation to predict the exceedance probability over the specified threshold at any location in the region of interest.

### 3.1. Estimation of temporal exceedance probabilities

In the initial step we model the temporal risks non-parametrically, by using the Nadaraya-Watson kernel estimator

$$\hat{\mathbf{P}}_{x_0}(t,s) = \frac{\sum_{i=1}^{n} K\left(\frac{t_i-t}{b_t}\right) 1_{\{X_s(t_i) \geq x_0\}}}{\sum_{i=1}^{n} K\left(\frac{t_i-t}{b_t}\right)}, \qquad (1)$$

where $K$ is a kernel function. For details on kernel smoothing we refer to Wand and Jones (1995). Note that the temporal bandwidth $b_t$ should not depend on the threshold $x_0$, in order for the resulting estimator to be non-increasing. In what follows, to keep notation simple, we write $b$ instead of $b_t$.

**Theorem 3.1.** *In the above notation, if $\frac{\partial^2}{\partial t^2}\left[\mathbf{P}_{x_0}(t,s)\right]$ exist a.e. in $[0,1]$ and if $\frac{\partial^2}{\partial t^2}\left[Var\left(1_{\{X_s(t)\geq x_0\}}\right)\right] < \infty$, as $n \to \infty$, $b \to 0$ and $nb \to \infty$, for all $s \in D$ and fixed $x_0 \in \mathbf{R}$, for the estimator (1) we have*

*(a) Consistency:*

$$MSE\left(\hat{\mathbf{P}}_{x_0}(t,s)\right) = O\left(max\left(b^4, \frac{1}{nb}\right)\right). \qquad (2)$$

*(b) Asymptotic normality:*

$$\frac{\hat{\mathbf{P}}_{x_0}(t,s) - E\hat{\mathbf{P}}_{x_0}(t,s)}{\sqrt{Var(\hat{\mathbf{P}}_{x_0}(t,s))}} \longrightarrow_d N(0,1). \qquad (3)$$

**Sketch of the proof.** By using Taylor expansions and integral approximations of the sums over the kernel function, it follows that $Bias(\hat{\mathbf{P}}_{x_0}(t,s)) = E\hat{\mathbf{P}}_{x_0}(t,s) - \mathbf{P}_{x_0}(t,s) = B(t,s,x_0)b^2 + o(b^2)$, where $B(t,s,x_0) = \frac{1}{2}\frac{\partial^2}{\partial t^2}\left[\mathbf{P}_{x_0}(t,s)\right] * int u^2 K(u) du$. Also, $Var(\hat{\mathbf{P}}_{x_0}(t,s)) = \frac{1}{(nb)^2}V(t,s,x_0) + o\left(\frac{1}{(nb)^2}\right)$, where $V(t,s,x_0)$ is bounded. Specifically, Ghosh and Draghicescu (2002) prove that there exist a temporal covariance function with the lag-k covariance denoted by $g(k,t,t')$ such that $V(t,s,x_0) = \sum_{i=1}^{n}\sum_{j=1}^{n} K\left(\frac{t_i-t}{b}\right) K\left(\frac{t_j-t'}{b}\right) g(|i-j|,t,t')$. Consistency then follows from Chebyshev's inequality. The asymptotic normality is an immediate application of results in Breuer and Major (1983).

**Remark 3.1.** Estimator (1) has the same rate of convergence as in the *iid* case.

**Remark 3.2.** An optimal bandwidth can be obtained by minimizing the mean squared error of the estimator (2). Note that, as in the *iid* case, $b_{opt} \sim Cn^{-\frac{1}{5}}$. In practice, an optimal bandwidth (local or global) can be obtained by using plug-in estimators (approximations) of the bias and variance (see for example Ghosh and Draghicescu 2002, Draghicescu 2003). While local (time-dependent) bandwidths have the advantage of dealing with edge effects (at the ends of the time interval), it is often the case that global (integrated over time) bandwidths considerably decrease computational time, without significant change in the resulting estimators.

**Remark 3.3.** In practice $Var(\hat{\mathbf{P}}_{x_0}(t,s))$ can be estimated by smoothing the empirical covariances, that provide good approximations for the above $g(k,t,t')$, and used to produce confidence bands based on the asymptotic normality of $\hat{\mathbf{P}}_{x_0}(t,s)$. In our preliminary analyses of the PM$_{10}$ data, this approach yielded very narrow bands.

### 3.2. Spatial interpolation

In the second step, we use use spatial interpolation (universal kriging) to predict the exceedance probability field at a location $s^* \in D$ where there are no observations, under the assumption of space-time separability. We model the spatial covariances $C_t(||s_i - s_j||) := Cov(\hat{\mathbf{P}}(t,s_i), \hat{\mathbf{P}}(t,s_j))$ parametrically, by using the Matèrn stationary covariance model

$$C_t(||s_i - s_j||) = \frac{\sigma_t}{2^{\nu_t-1}\Gamma(\nu_t)} \left(\frac{2\sqrt{\nu_t}||s_i - s_j||}{\rho_t}\right)^{\nu_t} \mathcal{K}_{\nu_t}\left(\frac{2\sqrt{\nu_t}||s_i - s_j||}{\rho_t}\right) \quad (4)$$

for fixed $t$. Our analyses did not detect major violations of spatial isotropy (that is, the spatial correlations only depend on distance, and not on direction). The threshold $x_0$ is also fixed. To keep notation simple, we omitted it in expression (4). $\Gamma(\cdot)$ is the usual gamma function and $\mathcal{K}_{\nu_t}(\cdot)$ is the modified Bessel function of the third kind of order $\nu_t$ (Abramowitz and Stegun 1972). The parameter $\nu_t > 0$ characterizes the smoothness of the process, $\sigma_t$ denotes the variance of the transformed random field, and $\rho_t$ measures how quickly the correlations of this field decay with distance. These parameters are estimated by maximum likelihood. Then, the best linear unbiased predictor (BLUP) of the exceedance probability field at $s_0 \in D$, is given by a linear combination $\hat{\mathbf{P}}^*(t, s_0) = \sum_{i=1}^{m} \lambda_i \hat{\mathbf{P}}(t, s_i)$, where the weights $\lambda_i$, $1 \leq i \leq m$ are completely determined by the covariance parameters $\nu_t, \rho_t$, and $\sigma_t$. The standard error of $\hat{\mathbf{P}}^*(t, s_0)$ can be also expressed in terms of the interpolation parameters $\lambda_i$. This procedure is known as *universal kriging* (Stein 1999), and implemented in many software packages. In practice, the covariance parameters are estimated from the same data. To account for their uncertainty, the standard errors of $\hat{\mathbf{P}}^*(t, s_0)$ need to be adjusted. This can be done by using conditional simulation techniques (Stein 1999, Chapter 6), or resampling schemes (Lahiri 2003).

**Remark 3.4.** Linear interpolation does not guarantee that the resulting exceedance probability estimator/predictor takes values in the interval $[0, 1]$. A $1:1$ transformation (such as $x \leftrightarrow \frac{e^x}{1+e^x}$) can be used first, then perform interpolation on the transformed field, and finally invert to obtain the desired exceedance probabilities. However, this technical detail did not provide improved maps for the present study.

### 4. Numerical simulations

In this section we present a simulation study that was carried out in order to analyze the performance of the exceedance probability estimator $\hat{\mathbf{P}}^*_{x_0}(t, s_0)$ introduced in Section 3. We simulated stationary isotropic space-time processes on a $20 \times 20$ grid and 200 time points. We considered a separable covariance model factorized as a Whittle-Matèrn spatial covariance by a stable temporal covariance, $C(u,h) = C_T(u)C_S(h) = \sigma_T^2 \exp(-u^\alpha) * \sigma_S^2 \frac{2^{1-\gamma}}{\Gamma(\gamma)} h^\gamma \mathcal{K}_\gamma(h)$, where $u = |t_l - t_k|$ is the time lag and $h = ||s_i - s_j||$ is the spatial distance between two sites, $\Gamma(\cdot)$ is the usual gamma function, and $\mathcal{K}_\gamma(\cdot)$ is the modified Bessel function of the third kind of order $\gamma$. The computations were done in R, using the function GaussRF in the library RandomFields. We used $\sigma_T^2 = 0.7$, $\sigma_S^2 = 1.3$, $\alpha = 0.2$, and $\gamma = 0.5$. One realization of the process is shown in Figure 4.

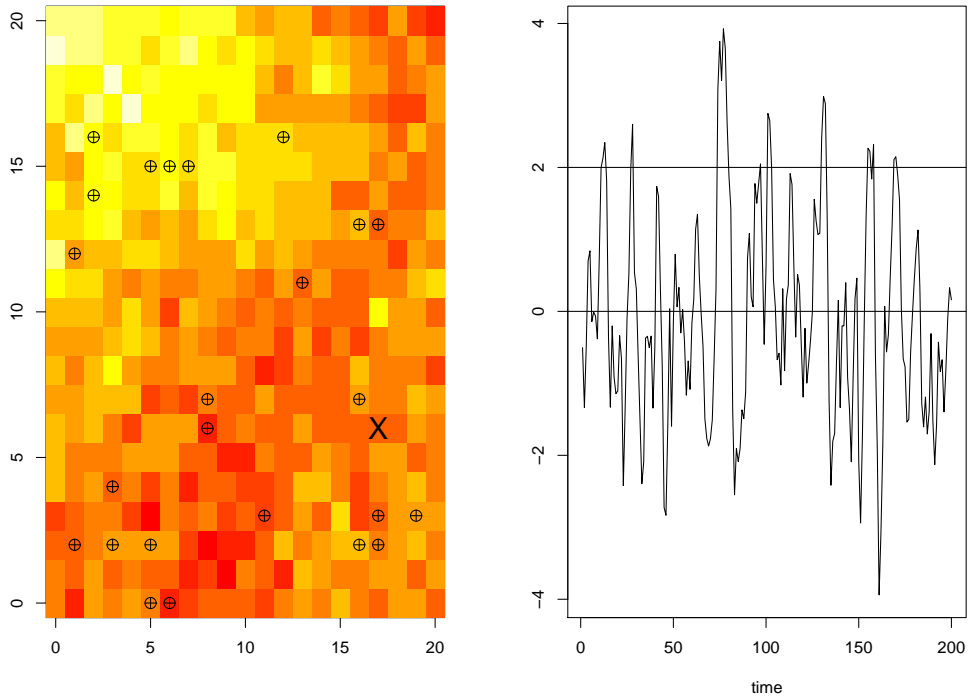

FIG 4. *One realization of the simulated spatial field at fixed time point $t = 100$ (left), superimposed locations of 25 randomly chosen sites; one realization of the process at location marked "X" (right).*

In order to assess the performance of the mapping procedure described in Section 3, we used the following three methods. They are all sequential, the second step being the same in all three cases (universal kriging with Matèrn covariance function). First method is indicator kriging, that is spatial interpolation of the empirical exceedance probabilities. It will be referred to as **IND.** The next approaches require a preliminary smoothing step in the temporal domain. Thus, the second method, referred to as **EDF**, is the two-step procedure introduced in Draghicescu and Ignaccolo (2005). For every time point $t_i$, a weight is assigned to the observed exceedance probability 0 or 1, corresponding to the order of the quantile of the $PM_{10}$ observation for that time point (i.e. the empirical distribution function EDF). For example, if the observed value is 80, and 75% of the data are less than 80, the respective weight will be 0.75, that is EDF(80). Then the exceedance probability is estimated by a weighted average of the observed exceedances on a time window centered at $t_i$. In the simulations, applications and validation study in next section, the window width was fixed and equal to 7 time points. The third method, called **KER** is based on kernel smoothing, as described in detail in Section 3.

TABLE 1
*Means and standard deviations over 100 simulations of root mean squared errors* $\sqrt{\sum_{i=1}^{200} \frac{(\hat{\mathbf{P}}^*_{x_0}(t_i,s_0)-\mathbf{P}^*_{x_0}(t_i,s_0))^2}{200}}$ *at location marked "X" in Figure 4, based on data at the randomly selected sites (m=24), and on data on the whole grid (m=400). Method* **IND**: *spatial interpolation of the observed 0 and 1 exceedance probabilities; method* **EDF**: *spatial interpolation of smoothed exceedance probabilities with weights proportional to the empirical distribution function; method* **KER**: *spatial interpolation of estimated exceedance probabilities obtained by kernel smoothing*

|  | $x_0 = 0$ | | $x_0 = 2$ | |
|---|---|---|---|---|
|  | $m = 24$ | $m = 400$ | $m = 24$ | $m = 400$ |
| **IND** | 0.772 (0.0698) | 0.0154 (0.1320) | 0.1001 (0.0759) | 0.078 (0.0521) |
| **EDF** | 0.0095 (0.0983) | 0.0098 (0.7095) | 0.0542 (0.0930) | 0.0327 (0.0198) |
| **KER** | 0.0083 (0.0572) | 0.0071 (0.0352) | 0.0493 (0.0087) | 0.0278 (0.0065) |

Table 1 shows the means and standard deviations over 100 simulations of the root mean squared errors $\sqrt{\sum_{i=1}^{200} \frac{(\hat{\mathbf{P}}^*_{x_0}(t_i,s_0)-\mathbf{P}^*_{x_0}(t_i,s_0))^2}{200}}$ at location marked "X" in Figure 4, for two thresholds, $x_0 = 0$ (corresponding to the median), and $x_0 = 2$ (corresponding to the 0.9 quantile). For each time point $t_i$, the predictor $\hat{\mathbf{P}}^*_{x_0}(t_i,s_0)$ was computed based on data at the randomly selected sites (m=24), as well as on data on the whole grid (m=400). It can be seen that for both thresholds these root mean squared errors are lowest for KER and largest for IND.

## 5. Maps of exceedance probabilities for $PM_{10}$ in Piemonte

We applied the three methods previously described to the Piemonte $PM_{10}$ data. We estimated the 50 $\mu g/m^3$ exceedance probabilities for each day of the year 2004.

Figure 5 displays these exceedance probabilities on February 18, 2004 (left), and February 19, 2004 (right), based on the aforementioned methods. On February 18 the maximum value of $PM_{10}$ for 2004 was observed at all sites, with a large decrease the following day. In this case IND does not seem to work well, the maps are very rough, and there is an abrupt change from a day to the next. The unshaded region on the February 19 map is due to negative values of the predicted exceedance probabilities, that, for the applied goal of this study, can be viewed as zeros. EDF yields maps with high values on almost all the region, while KER shows higher values around the Torino area. As observations outside

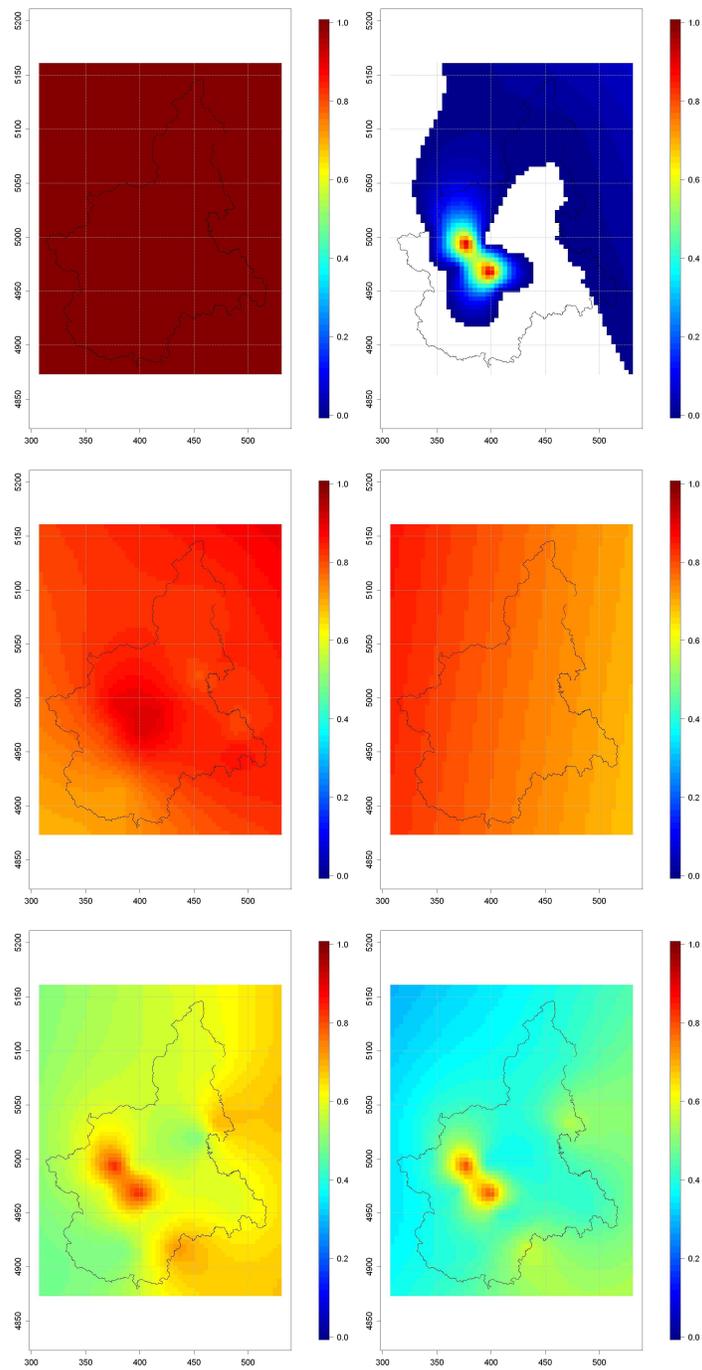

Fig 5. *Estimated $50\mu g/m^3$ exceedance probabilities on February 18, 2004 (left) and February 19, 2004 (right); IND (top), EDF (middle), KER (bottom).*

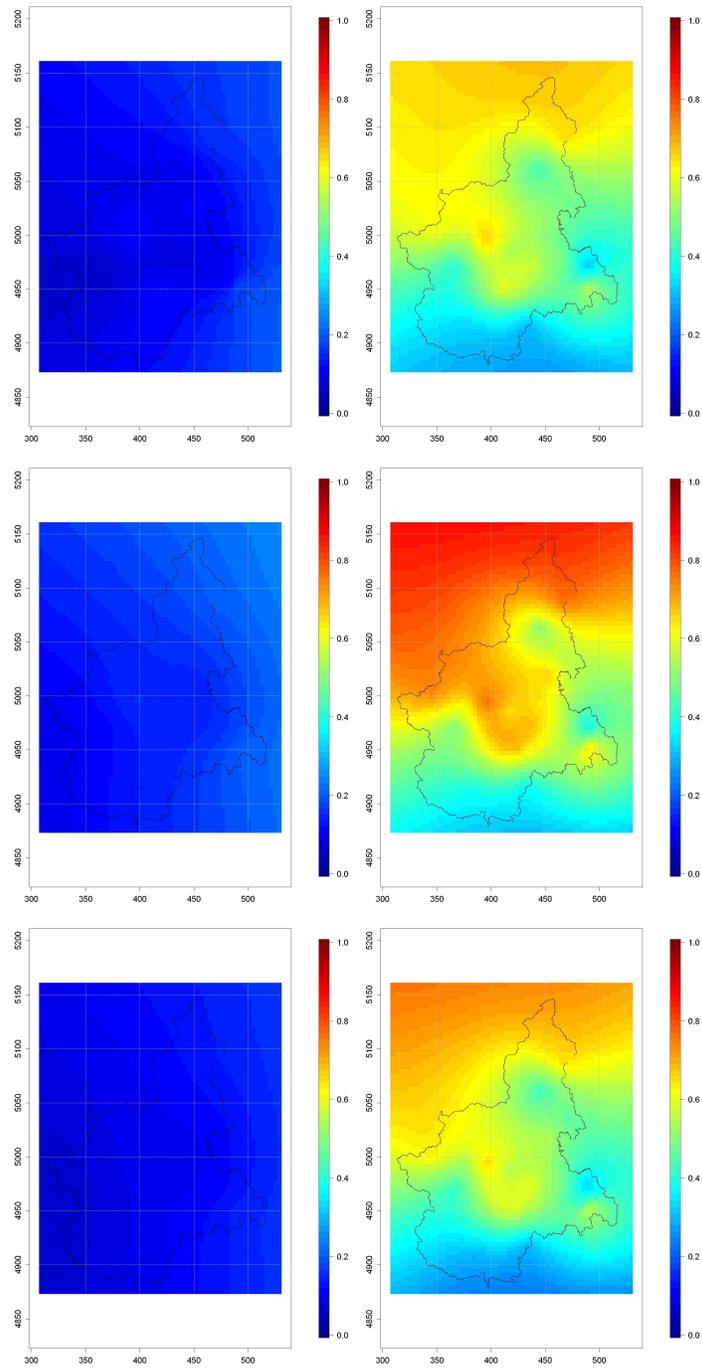

Fig 6. *Estimated $50\mu g/m^3$ exceedance probabilities averaged over summer (left) and winter (right); IND (top), EDF (middle), KER (bottom).*

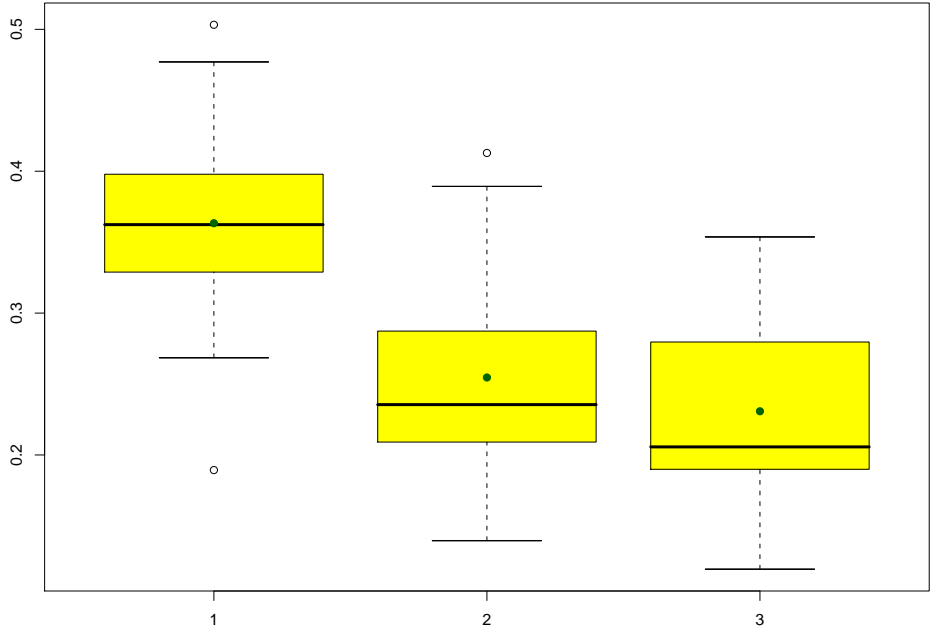

Fig 7. *Boxplots (over the 22 locations) of the root mean squaerd prediction errors for 50 $\mu g/m^3$ exceeding probabilities: IND (left), EDF (middle), KER (right).*

Piemonte were not available, the estimated exceedance probabilities near the boundaries tend to have larger errors.

While these methods provide maps of point estimates, and could give a good understanding of the process (by using animation techniques for example), it is often the case that transformations of point estimates are also of interest. For instance, seasonal averages are very informative summaries for both scientists and policy makers. Figure 6 shows estimated exceedance probabilities over 50 $\mu g/m^3$ averaged over summer (left) and winter (right), obtained through the three methods. Summer is considered the period between April 1 and September 30. From a geographical prospective these methods yield maps that are not very different. However, from a statistical point of view, EDF or KER have the advantage of making use of all the information in the data. As expected, in both seasons the highest risks are around the Torino area, with larger values during the winter.

To assess the performance of these methods, we carried out a cross-validation study. We used the leave-one-out principle and estimated the daily 50 $\mu g/m^3$ exceedance probabilities at each site, based on observations at the remaining 21 sites for each day. The boxplots in Figure 7 display the root mean squared prediction errors obtained by leaving out that site, computed as the square root of the averages (over the 366 days of the year 2004) of the differences in squared exceedance probabilities ("observed" - predicted). Here, the "observed" values are 0 or 1 for IND, and the corresponding smoothed values (after first step) for EDF and KER, respectively. As expected, IND performs the worst, while KER yields the smallest errors.

## 6. Discussion

The methodology proposed in this paper provides a good descriptive and visual tool for modeling threshold exceedance probabilities based on space-time data with large temporal and relatively small spatial coverage. It is statistically accurate, computationally fast, and flexible enough to be suitable for processes with complex space-time dependencies in many applied fields, such as environmental science and management, atmospheric sciences, ecology, epidemiology, finance, medicine. The method KER, involving kernel smoothing in time, followed by spatial interpolation, was proved to provide the most accurate and informative maps of exceedance probabilities.

## Acknowledgements

The authors kindly thank Giorgio Arduino, who provided useful information and access to data, and Stefano Bande, who helped with the geocoding for the air pollution maps. We are also grateful to the Editor, Associate Editor, and two reviewers for their suggestions that helped improve the presentation of this work. The work of R. Ignaccolo was partially supported by the PRIN project n.2006131039 and the Regione Piemonte CIPE project 2004.

Here:

...